\documentclass[prx,aps,superscriptaddress,nofootinbib,twocolumn]{revtex4-1}

\usepackage{mathrsfs}
\usepackage{amsfonts}
\usepackage{amssymb}
\usepackage{amsmath}
\usepackage{graphicx}
\usepackage[usenames,dvipsnames]{color}
\usepackage[colorlinks=true,citecolor=blue,linkcolor=magenta]{hyperref}
\usepackage{ulem}
\usepackage{lmodern}

\newcommand{\figpath}{.}

\renewcommand{\Re}{\mathrm{Re}}

\newcommand{\Tr}{\mathrm{Tr}}
\newcommand{\norm}[1]{\Vert #1 \Vert}
\newcommand{\abs}[1]{\vert #1 \vert}

\newcommand{\ket}[1]{\vert{ #1 }\rangle}
\newcommand{\bra}[1]{\langle{ #1 }\vert}
\newcommand{\ketbra}[2]{\vert #1 \rangle \langle #2 \vert}
\newcommand{\braket}[2]{\langle #1 \vert #2 \rangle}
\newcommand{\mean}[1]{\langle #1 \rangle}

\newcommand{\HC}{{\rm h.c.}}

\newcommand{\SIG}[2]{\sigma^{#1}_{#2}}

\begin{document}

\title{Efficient variational quantum simulator incorporating active error minimisation}
\author{Ying Li}

\affiliation{Department of Materials, University of Oxford, Parks Road, Oxford OX1 3PH, United Kingdom}

\author{Simon C. Benjamin}

\affiliation{Department of Materials, University of Oxford, Parks Road, Oxford OX1 3PH, United Kingdom}

\begin{abstract}
One of the key applications for quantum computers will be the simulation of other quantum systems that arise in chemistry, materials science, etc, in order to accelerate the process of discovery. It is important to ask the following question: Can this simulation be achieved using near future quantum processors, of modest size and under imperfect control, or must it await the more distant era of large-scale fault-tolerant quantum computing? Here we propose a variational method involving closely integrated classical and quantum  coprocessors. We presume that all operations in the quantum coprocessor are prone to error. The impact of such errors is minimised by {\it boosting} them artificially and then extrapolating to the zero-error case. In comparison to a more conventional optimised Trotterisation technique, we find that our protocol is efficient and appears to be fundamentally more robust against error accumulation. 
\end{abstract}

\maketitle

\section{Introduction}

Many quantum algorithms have been developed under the presumption that the hardware upon which they will run is effectively error-free: the error rate is so low that the entire algorithm can be executed successfully without a single error. It is now known that such hardware can, in principle, be created using components that have far higher error rates. {\it Fault-tolerant} quantum computing can be achieved by encoding qubits in non-Abelian anyons in topological materials~\cite{Kitaev2003} or using the quantum error correction codes~\cite{Raussendorf2007}. While the former is still in the early stages of its development, for the latter approach sub-threshold quantum operations have been demonstrated in ion-trap and superconducting systems~\cite{Barends2014, Harty2014, Lucas, Wineland}. However, quantum error correction involves a substantial multiplication of resources; the number of physical qubits required may be orders of magnitude greater than the number of error-free logical qubits seen by the algorithm. A recent study audited the cost of implementing Shor's algorithm to solve a classically-infeasible task, and found that even with state-of-the-art techniques for magic state distillation the machine would need over six million of today's highest quality qubits~\cite{Joe2016}. 

The need for millions of qubits contrasts starkly with the fact that only fifty qubits are needed to achieve so-called `quantum supremacy', i.e. to create a quantum processor that is so complex that conventional supercomputers~\cite{qHiPSTER} cannot predict its behaviour. Machines involving this many qubits, under good but imperfect control, are expected to emerge in the next few years. The challenge for researchers is to identify useful functions for such devices, in order to motivate further investment and continue the evolution toward the longer-term goal of fully fault tolerant systems. 

Recently some hybrid quantum/classical algorithms have been developed which are promising for near future quantum applications~\cite{Farhi2014, Peruzzo2014, Wecker2015, McClean2016, Bauer2016, Kreula2016, Kreula2016EPJ, Shen2015, OMalley2016}. A common feature of these algorithms is that the quantum computer is only in charge of carrying out a subroutine, acting as a `coprocessor' while the larger scale algorithm is governed by a classical computer. The task of the quantum computer is thus simplified and may be accomplished with relatively few quantum operations. A higher error rate per operation is therefore tolerable; in a fault-tolerant machine this would imply a more modest resource overhead for the code, but it may even be possible to implement such quantum algorithms without quantum error correction. 

Hybrid approaches are very relevant to quantum simulation, i.e. Feynman's vision~\cite{Feynman1982,Lloyd1996} of using a controlled quantum processor to model another quantum system. Such a technology would be highly advantageous for the investigation of various large quantum systems, e.g.~simulating quantum chemistry systems~\cite{Kassal2011,Wecker2014, Poulin2015, Reiher2016}, or novel materials and other condensed-matter systems~\cite{Lanyon2011, Ma2011, Barends2015}. A powerful tool that has been exploited in several hybrid protocols is the variational method~\cite{Farhi2014, Peruzzo2014, Wecker2015, McClean2016}. Typically the state of the target system can be found by writing a trial quantum state with a large but tractable number of parameters, and then discovering the optimal value of these parameters. Implicitly this requires the scientists to use their understanding of the target system (the novel molecule, or material) to select a set of parameters that, while large, is far smaller than the total number of parameters needed to specify an arbitrary quantum state. The latter is of course exponential in the number of particles composing the target system. 

\begin{figure}[tbp]
\centering
\includegraphics[width=1\linewidth]{\figpath 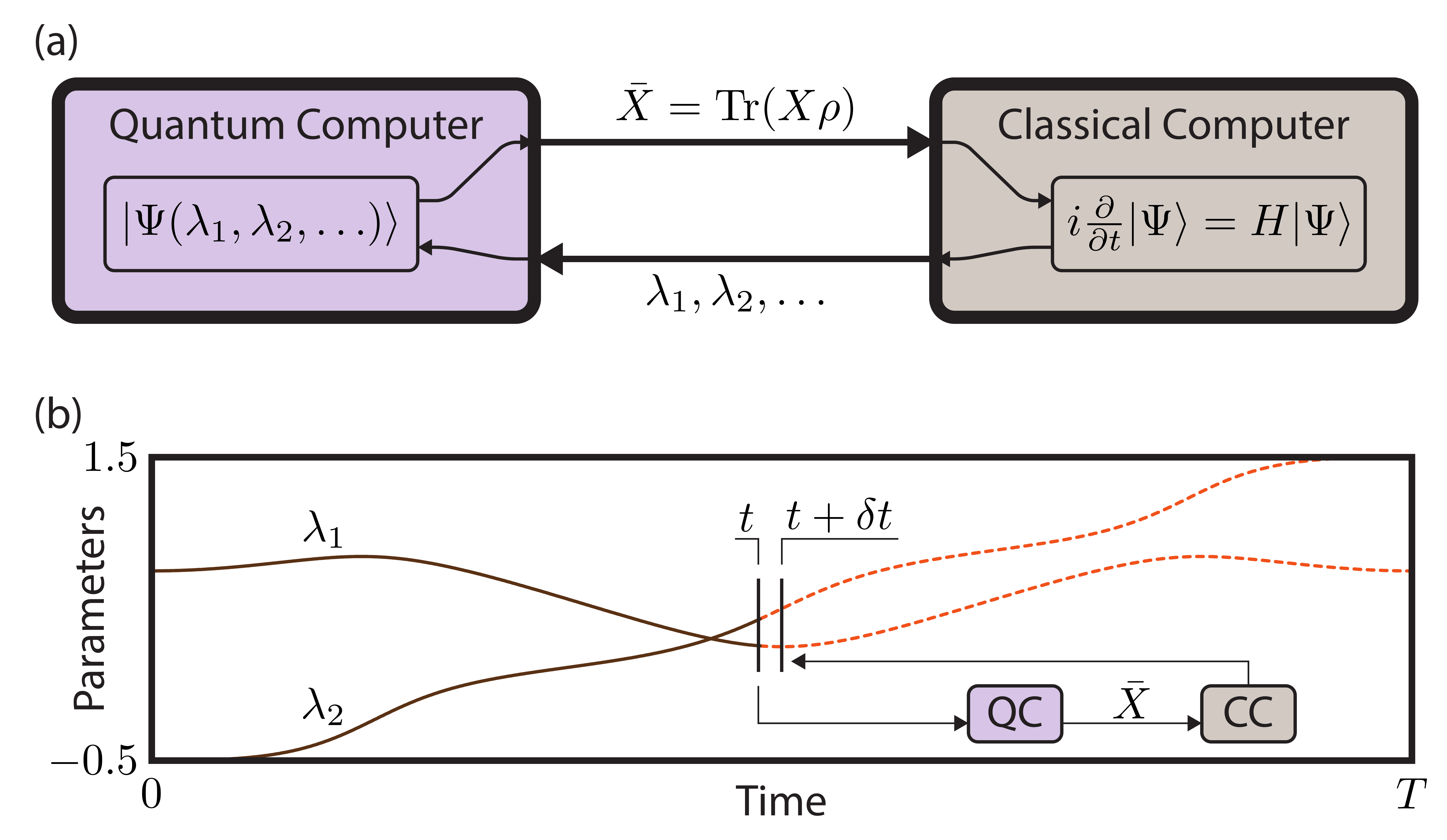}
\caption{
Hybrid solver of quantum dynamics. (a) Both a quantum computer and a classical computer are used in solving the time evolution of a quantum system. The quantum state is approximated by a trial state $\ket{\Psi(\lambda _{1},\lambda _{2},\ldots)}$. Variational parameters are determined by the classical computer according to the Schr\"{o}dinger equation. The quantum computer is used to implement a subroutine: Inputs are parameters $\{\lambda _{k}(t)\}$, and outputs are values of certain derivatives required by the main program in the classical computer. (b) Variational parameters are determined iteratively given their initial values. Parameters at the time $t$ are sent to the quantum computer, which evaluates the quantities required by the classical computer. Then, the classical computer can work out parameters at the time $t+\delta t$, where $\delta t$ is a short time. Note that the curves here represent the actual evolution of parameters in the example described in Appendix~\ref{sec:model}.
}
\label{fig:hybrid}
\end{figure}

Our focus here is on finding the dynamics of interesting quantum systems, and we briefly remark on the considerable significance of such a capability. Dynamics must be studied when properties cannot be determined from static features. This has motivated dynamical versions of many well-known techniques, e.g.~nonequilibrium dynamical mean-field theory~\cite{Aoki2014}, the time-dependent variational quantum Monte Carlo method~\cite{Carleo2012}, time-dependent tensor network methods~\cite{Daley2004, Banuls2009}, and of course time-dependent density functional theory~\cite{Runge1984}. However, there are still many problems that cannot be solved using these powerful classical methods, so it is hoped that quantum computers can extend their reach~\cite{Feynman1982, Lloyd1996, Cirac2012}.

We therefore propose a hybrid quantum algorithm for simulating the dynamics of a quantum system. The conventional approach for simulating quantum dynamics employs Trotterisation~\cite{Lloyd1996, Trotter1959, Ortiz2001, Dhand2014, Knee2015}, which usually requires many quantum operations; therefore it seems likely to necessitate the full machinery of fault-tolerant quantum computing~\cite{Wecker2014, Poulin2015, Reiher2016}. Our approach is based on the variational method and our hope is that it could be implemented using small-size quantum circuits, i.e.~quantum circuits with a small number of quantum operations that suffer significant noise compared with fault-tolerant quantum computers. A novel feature of our algorithm is that it compensates for errors through classical inference without encoding: If the noise in the quantum computer mainly results in stochastic errors and the rate of errors can be amplified in a controllable way, then we find that errors can be approximately corrected. The condition that noise is stochastic can be met by engineering for many systems: if, for example, single qubit gates have relatively high fidelity~\cite{Barends2014, Harty2014, Lucas, Wineland, Ladd2010} and their noise is stochastic, then arbitrary two-qubit gate errors can be made stochastic through a twirling-like technique~\cite{Knill2004, Wallman2015, OGorman2016} which we presently discuss.  Moreover, the severity of such errors can be deliberately {\it increased} artificially, allowing one to create curves that the classical algorithm can extrapolate to estimate the zero-error limit. We performed numerical emulations of the process on small systems, finding that this technique does indeed lead to robustness: The impact of physical errors on the simulator's performance is far lower than in an (optimised) Trotterisation protocol, and moreover this impact does not worsen with the duration of the simulation.

The remainder of this paper is organised as follows. In Sec.~\ref{sec:review}, we review the Trotterisation algorithm and variational methods. In Sec.~\ref{sec:algorithm}, our hybrid algorithm is introduced. In Sec.~\ref{sec:principle}, the variational theory is discussed. In Sec.~\ref{sec:hybrid}, the task for the quantum computer and the overall program are described in detail. Errors in our algorithm are analysed in Sec.~\ref{sec:analysis}. The method for reducing errors is given in Sec.~\ref{sec:noise}, in which we also discuss how to convert non-stochastic errors into stochastic errors and how to tune the rate of errors. Numerical results are presented in Sec.~\ref{sec:numerics}. A summary is given in Sec.~\ref{sec:summary}.

\section{Trotterisation and variational method}
\label{sec:review}

The Trotterisation approach to simulation, which we use as a basis for comparison with our protocol, exploits the fact that time evolution under a general Hamiltonian $H = \sum _j H_j$ can be approximated according to the Trotter-Suzuki decomposition~\cite{Trotter1959}
\begin{eqnarray}
e^{-iHT} \simeq R = \prod _{n=1}^{N_{\rm t}} \left( \prod _j e^{-iH_j \tau_{n,j}} \right).
\label{eq:decomposition}
\end{eqnarray}
Here, each term $e^{-iH_j \tau_{n,j}}$ corresponds to the evolution driven by the term $H_j$ for a short time $\tau_{n,j}$, which can be realised by a quantum gate or a combination of quantum gates. Usually, the short time is taken uniformly as $\tau_{n,j} = T/N_{\rm t}$, where $T$ is the time of the simulated evolution. When $N_{\rm t}$ is larger, the approximation is better, and errors in the approximation scale with the simulated time and the number of quantum gates as $T^2 / N_{\rm t}$~\cite{Dhand2014}.

Our approach is based on a variational technique. Variational methods have numerous applications in the numerical study of many-body quantum systems, for examples, density functional theory~\cite{Jones2015}, the matrix product state method~\cite{PerezGarcia2007}, and simulating molecular dynamics using the variational principle~\cite{Feldmeier2000}. In these methods, typically a trial state is used to approximate the true state of the system. The trial state must of course be specified by some tractable number of parameters. But since existing realisations are entirely classical, there is a stronger condition on the trial function: it must be possible to efficiently evaluate its fit to the true quantum state using only a classical algorithm. This requirement limits the application of variational methods. Sometimes it may be impossible to evaluate a trial state that provides a good approximation to the true state in a classical computer. In such a case, a quantum computer could be helpful, because we may be able to complete tasks that are difficult for a classical computer using a quantum computer. An example is the unitary coupled cluster method~\cite{Peruzzo2014, Taube2006}, in which the energy of the trial state can be evaluated using a quantum computer when it is hard for a classical computer to do so. The protocol we describe here is another example.

\section{Hybrid quantum simulation of dynamics}
\label{sec:algorithm}

\begin{figure*}[tbp]
\centering
\includegraphics[width=0.65\linewidth]{\figpath 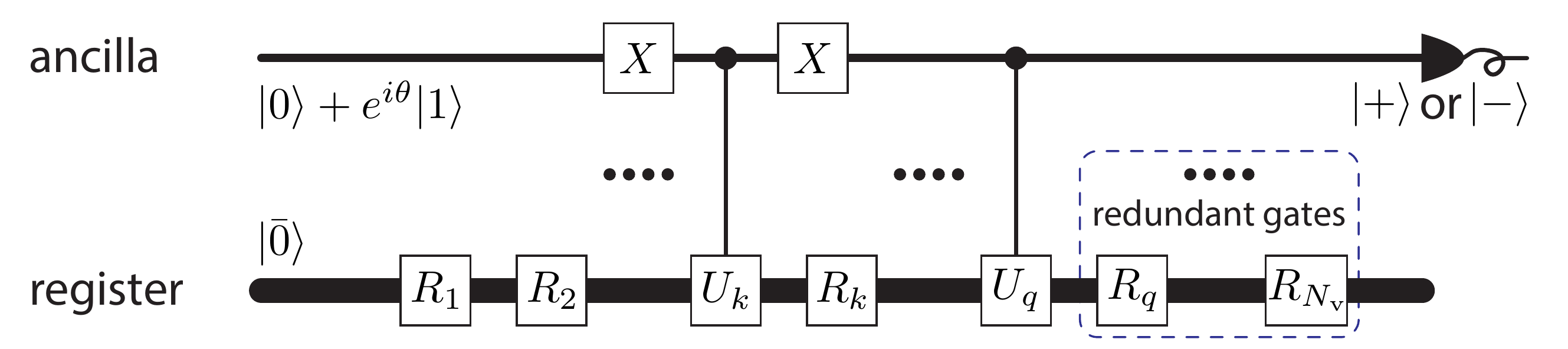}
\caption{
Quantum circuit for the evaluation of certain coefficients required by the classical main program, as specified in the text. To evaluate $\Re \left( e^{i\theta} \bra{\bar{0}} U \ket{\bar{0}} \right)$, where $U = R_{1}^\dag \cdots U_{k}^\dag R_{k}^\dag \cdots R_{N_{\rm v}}^\dag R_{N_{\rm v}} \cdots R_{q} U_{q} \cdots R_{1}$, the ancillary qubit is initialised in the state $(\ket{0}+e^{i\theta}\ket{1})/\sqrt{2}$ and measured in the $\ket{\pm} = (\ket{0} \pm \ket{1})/\sqrt{2}$ basis. Here, $U_{k}$ is one of $\sigma_{k,i}$, and $U_{q}$ is one of $\sigma_{q,j}$ or $\sigma_{j}$ (By taking $q = N_{\rm v}+1$, $\sigma_{j}$ is put on the left side of $R_{N_{\rm v}}$ in the product). In the figure, we have assumed that $k < q$. Gates on the register after the second controlled unitary gate can be omitted. This circuit is actually a variant of the circuit proposed in 2002 by Ekert {\it et al}~\cite{Ekert2002, footnote2}. It involves $N_{\rm v}$ gates on the register, two flip gates ($X$) on the ancillary qubit, and two controlled unitary gates on the ancillary qubit and the register.
}
\label{fig:circuit}
\end{figure*}

The purpose of our hybrid algorithm is to solve the Schr\"{o}dinger equation $i \frac{\partial}{\partial t} \ket{\Phi(t)} = H \ket{\Phi(t)}$ ($\hbar = 1$), assuming that the state $\ket{\Phi(t)}$ can be approximated by a trial state $\ket{\Psi(t)} \equiv \ket{\Psi(\lambda_{1},\lambda_{2},\ldots)}$, where $\{\lambda _{k}(t)\}$ are variational parameters. As shown in Fig.~\ref{fig:hybrid}(a), the hybrid algorithm is implemented on both a quantum computer and a classical computer. The task of the classical computer is to determine variational parameters according to the Schr\"{o}dinger equation, and this procedure requires certain derivatives associated with the state $\ket{\Psi(t)}$ which the quantum computer provides.

The hybrid algorithm works out variational parameters iteratively as shown in Fig.~\ref{fig:hybrid}(b). Parameters at the time $t$ ($\{\lambda _{k}(t)\}$) are sent to the quantum computer, with which the quantum computer finds the values required by the classical computer. Based on results from the quantum computer, the classical computer can determine parameters at the time $t+\delta t$ ($\{\lambda _{k}(t+\delta t)\}$), where $\delta t$ is a short time. Then, these new parameters are sent back to the quantum computer. In this way, given parameters of the initial state ($\{\lambda _{k}(0)\}$), parameters at the time $T$ ($\{\lambda _{k}(T)\}$) are systematically inferred by iterating the process carried out by two computers. The simulation is successful if the state $\ket{\Psi(T)}$ is a good approximation of the state $\ket{\Phi(T)}$.

Using the variational method, the degrees of freedom provided by variational parameters allow us to use quantum circuits with a size much smaller than the circuit of the Trotterisation algorithm to simulate the time evolution of a quantum system. Note that this is an `apples to oranges' comparison because our algorithm only simulates the time evolution of a given initial state while the Trotterisation algorithm simulates the time evolution of arbitrary initial states, i.e.~the time evolution operator. Thus our algorithm aims at an easier problem than the Trotterisation algorithm. 

Tracking the evolution from a specific initial state is the main goal in many simulations, and other more general tasks can also be reexpressed this way. The approach we describe can be relevant to the specific problem of designing and calibrating quantum gates, thus allowing early quantum computers to aid in the design of their successors. Moreover there are also interesting connections between dynamical simulation and the determination of static properties: one could find a ground state by simulating an adiabatic time evolution~\cite{Farhi2000}, thus our algorithm is relevant to that task. In other hybrid algorithms for determining the ground state of a quantum system~\cite{Farhi2014, Peruzzo2014, Wecker2015, McClean2016}, one may need to find the global minimum of the energy in the parameter space to maximise the fidelity. However, finding the global minimum in a high-dimensional parameter space is usually a non-trivial computing task. In our algorithm, parameters are worked out iteratively, therefore the global minimisation is not required. We remark that Trotterisation is used in some hybrid algorithms~\cite{Peruzzo2014, McClean2016, Bauer2016, Kreula2016, Kreula2016EPJ}. In principle our algorithm can be used to replace the Trotterisation method in these instances, to further simplify the task of the quantum computer.

\section{Variational theory of quantum time evolution}
\label{sec:principle}

The time-dependent variational principle corresponding to the Schr\"{o}dinger equation reads $\delta \int_{t_i}^{t_f} dt L = 0$, where the Lagrangian is~\cite{Dirac1930, Frenkel1934}
\begin{eqnarray}
L = \bra{\Psi(t)} (i\frac{\partial}{\partial t} - H) \ket{\Psi(t)}.
\end{eqnarray}
Assuming that the state $\ket{\Psi(t)}$ is determined by a set of real parameters $\{\lambda_{k}(t)\}$, i.e.~$\ket{\Psi(t)} \equiv \ket{\Psi(\lambda_{1},\lambda_{2},\ldots)}$, the Lagrangian can be rewritten as
\begin{eqnarray}
L = i\sum_{k} \bra{\Psi} \frac{\partial \ket{\Psi}}{\partial \lambda_{k}} \dot{\lambda}_{k}
- \bra{\Psi} H \ket{\Psi},
\end{eqnarray}
which is a function of parameters $\{\lambda_{k}\}$ and their time derivatives $\{\dot{\lambda}_{k} = \frac{d\lambda_{k}}{dt}\}$. According to $L$, the Euler-Lagrange equation describing the evolution of parameters (hence the state $\ket{\Psi}$) is
\begin{eqnarray}
\sum_{q} M_{k,q} \dot{\lambda}_{q} = V_{k},
\label{eq:DifEq}
\end{eqnarray}
where
\begin{eqnarray}
M_{k,q} &=& 
i \eta \frac{\partial \bra{\Psi}}{\partial \lambda_{k}}
\frac{\partial \ket{\Psi}}{\partial \lambda_{q}}
+\HC , \\
V_{k} &=&
\eta \frac{\partial \bra{\Psi}}{\partial \lambda_{k}} H \ket{\Psi}
+\HC .
\end{eqnarray}
Here, $\eta = 1$, both $M$ and $V$ are real, and $M$ is antisymmetric. There are other variational principles for the quantum time evolution~\cite{Broeckhove1988}. For example, McLachlan's variational principle reads $\delta \Vert (i\frac{\partial}{\partial t} - H) \ket{\Psi(t)} \Vert = 0$~\cite{McLachlan1964}, which leads to the same equation as Eq.~(\ref{eq:DifEq}) but $\eta = -i$. Here, the norm is $\norm{\psi} = \sqrt{\braket{\psi}{\psi}}$. In the following, we focus on the time-dependent variational principle, but the hybrid algorithm can be adapted to McLachlan's variational principle.

Recall that we can always express a state as $\ket{\Psi} = \sum_{n}(\alpha_{n}+i\beta_{n})\ket{n}$, where $\alpha_{n}$ and $\beta{_n}$ are real, and $\ket{n}$ are orthonormal basis states. Taking parameters $\{\lambda_{k}\} = \{\alpha_{n},\beta_{n}\}$, Eq.~(\ref{eq:DifEq}) leads to the Schr\"{o}dinger equation; but of course we require a parameterisation such that the number of parameters remains tractable for the sizes of target systems that we are interested in. Thus our variational approach, like others, is relevant when the scientist can make an educated guess as to the general form of the quantum state -- there can be a large number of free parameters $\{\lambda_{k}\} $, but typically far fewer than would be needed to specify an arbitrary state.

\section{Variational algorithm on a hybrid computer}
\label{sec:hybrid}

We consider trial states that can be directly prepared in the quantum computer, i.e.~states can be expressed as $\ket{\Psi} = R\ket{\bar{0}}$, where $\ket{\bar{0}}$ is an initial state of the quantum computer, and $R$ is a sequence of quantum gates determined by parameters $\{\lambda_{k}\}$, i.e.~
\begin{eqnarray}
R = R_{N_{\rm v}}(\lambda_{N_{\rm v}}) R_{N_{\rm v}-1}(\lambda_{N_{\rm v}-1}) \cdots R_{2}(\lambda_{2}) R_{1}(\lambda_{1}).
\end{eqnarray}
Here, $R_k$ is a unitary operator describing a quantum gate, and the total number of gates (i.e.~parameters) is $N_{\rm v}$. If $N_{\rm v}$ is smaller than the dimension of the Hilbert space ($2\times {\rm dim}-2$ to be exact), trial states $\ket{\Psi}$ only span a sub-manifold of the Hilbert space. In this restricted trial-state space, Eq.~(\ref{eq:DifEq}) approximates the exact evolution if the exact state is close to the trial-state space.

In the following analysis, we describe each $R_k$ gate as dependent on only one parameter $\lambda_k$. However it is worth noting that the trial state can be generalised to the case where each gate $R_k$ depends on multiple parameters, including parameters that vary in a pre-defined way with time, and that both the Trotter-Suzuki decomposition~\cite{Trotter1959} and the unitary coupled cluster ansatz~\cite{Taube2006} can be expressed in this form. As a generalisation to the case in which the number of gates $N_{\rm v}$ is fixed, one can even vary $N_{\rm v}$ depending on the simulated time, providing that we understand how to re-express the trial state using the new gates (adding gates to the set is of course trivially possible). Our point is that the set of gates with which we create our trial state, can itself evolve over the simulated time.

We are interested in the case where evaluating coefficients $M$ and $V$ in Eq.~(\ref{eq:DifEq}) is intractable in a classical computer, therefore these coefficients are obtained using the quantum computer. Each parameter is determined in turn, by appropriately configuring a quantum circuit involving $\sim N_{\rm v}$ gates and a single measurement outcome; this fixed circuit is run repeatedly until the expected measurement outcome is known to a given precision. Note that this implies the overall protocol is trivially parallelisable over a large number of quantum processors with no quantum link between them.

We express the Hamiltonian in the form
\begin{eqnarray}
H = \sum_{i} h_i \sigma_{i},
\label{eq:Ham}
\end{eqnarray}
where $\sigma_{i}$ are unitary operators. In many quantum systems, the number of terms in this expression scales with the size of the system polynomially. Similarly, we write
\begin{eqnarray}
\frac{d R_{k}}{d \lambda_{k}} = \sum_{i} f_{k,i} R_{k} \sigma_{k,i},
\label{eq:dRk}
\end{eqnarray}
where $\sigma_{k,i}$ are also unitary operators. For many frequently used single-qubit gates and two-qubit gates ($R_{k}$), there is only one term in this expression, and $\sigma_{k,i}$ is also a one-qubit or two-qubit gate. Because any operator can be expressed using Pauli operators, we can choose unitary operators $\sigma_{i}$ and $\sigma_{k,i}$ as (single-qubit and multi-qubit) Pauli operators.

Using the expression (\ref{eq:dRk}), we rewrite the derivative of the state as
\begin{eqnarray}
\frac{\partial \ket{\Psi}}{\partial \lambda_{k}} = \sum_{i} f_{k,i} R_{k,i} \ket{\bar{0}},
\end{eqnarray}
where
\begin{eqnarray}
R_{k,i}= R_{N_{\rm v}} R_{N_{\rm v}-1} \cdots R_{k+1} R_{k} \sigma_{k,i} \cdots R_{2} R_{1}.
\end{eqnarray}
Then, differential equation coefficients can be expressed as
\begin{eqnarray}
M_{k,q} = \sum_{i,j} \left(
i f^*_{k,i}f_{q,j} \bra{\bar{0}} R^\dag_{k,i} R_{q,j} \ket{\bar{0}}
+ \HC \right),
\label{eq:M}
\end{eqnarray}
and
\begin{eqnarray}
V_{k} = \sum_{i,j} \left(
f^*_{k,i}h_{j} \bra{\bar{0}} R^\dag_{k,i}\sigma_{j}R \ket{\bar{0}}
+ \HC \right),
\label{eq:V}
\end{eqnarray}
where we have used the expression (\ref{eq:Ham}).

In Eqs.~(\ref{eq:M})~and~(\ref{eq:V}), each term is in the form
$$
a \Re \left( e^{i\theta} \bra{\bar{0}} U \ket{\bar{0}} \right),
$$
where the amplitude $a$ and phase $\theta$ are determined by either $i f^*_{k,i}f_{q,j}$ or $f^*_{k,i}h_{j}$~\cite{footnote1}, and $U$ is a unitary operator equal to either $R^\dag_{k,i}R_{q,j}$ or $R^\dag_{k,i}\sigma_{j}R$. Such a term can be evaluated using the quantum circuit shown in Fig.~\ref{fig:circuit}. This circuit needs an ancillary qubit initialised in the state $(\ket{0}+e^{i\theta}\ket{1})/\sqrt{2}$ and a register initialised in the state $\ket{\bar{0}}$. The ancillary qubit is measured in the $\{\ket{+},\ket{-}\}$ basis after a sequence of quantum gates on the register and two controlled gates, in which the ancillary qubit is the control qubit. The value of each term is given by $\Re \left( e^{i\theta} \bra{\bar{0}} U \ket{\bar{0}} \right) = \mean{X} = \Tr(X\rho)$, where $\rho$ is the final state of the quantum computer, and $X$ is the x-direction Pauli operator of the ancillary qubit. In the following, we consider the case in which the value of $\mean{X}$ is estimated by repeating this relatively shallow circuit and calculating the mean value of measurement outcomes. Note that the value of $\mean{X}$ could be estimated more efficiently using quantum amplitude estimation~\cite{Brassard2002} if error rates could be made low enough to allow a circuit of sufficient depth to function.

\subsection*{Main program}

The overall flow of the algorithm is as follows: Firstly, we select initial parameters $\{\lambda_{k}(0)\}$. Secondly, we solve the differential equation~(\ref{eq:DifEq}) numerically using the classical computer, in which the matrix $M$ and the vector $V$ in the equation are evaluated using the quantum coprocessor. The solution permits us to project our parameters forward by a small time increment, and repeat the second step. Eventually we reach the parameters $\{\lambda_{k}(T)\}$ which allow us to prepare the final state in the quantum computer. 

There are many different numerical methods for solving a differential equation, and the choice of specific numerical method determines the details of the information exchange loop between quantum and classical processors. In the following, we take the Euler method as an example, but the algorithm can be adapted to other numerical methods, e.g.~Runge-Kutta methods~\cite{Butcher2008}.

Time is discretised as $t_n = n\delta t$, where $t_0 = 0$ is the initial time, and $t_N = N\delta t = T$ is the simulated evolution time. Firstly, $M$ and $V$ corresponding to parameters $\{\lambda_k(t_0)\}$ are evaluated using the quantum computer. Then, the following process is repeated. Given $M$ and $V$ corresponding to the time $t_n$, Eq.~(\ref{eq:DifEq}) is solved numerically on the classical computer to obtain values of $\{\dot{\lambda}_{k}(t_n)\}$. As parameters $\{\lambda_k(t_n)\}$ have been obtained from previous calculations, one can approximately calculate parameters of the time $t_{n+1}$ using $\lambda_k(t_{n+1}) = \lambda_k(t_{n}) + \dot{\lambda}_{k} \delta t$. Repeating the process until $t_{n+1} = T$, we can work out the parameters $\{\lambda_{k}(T)\}$ of the final state.

\section{Error analysis}
\label{sec:analysis}

\begin{figure}[tbp]
\centering
\includegraphics[width=1\linewidth]{\figpath 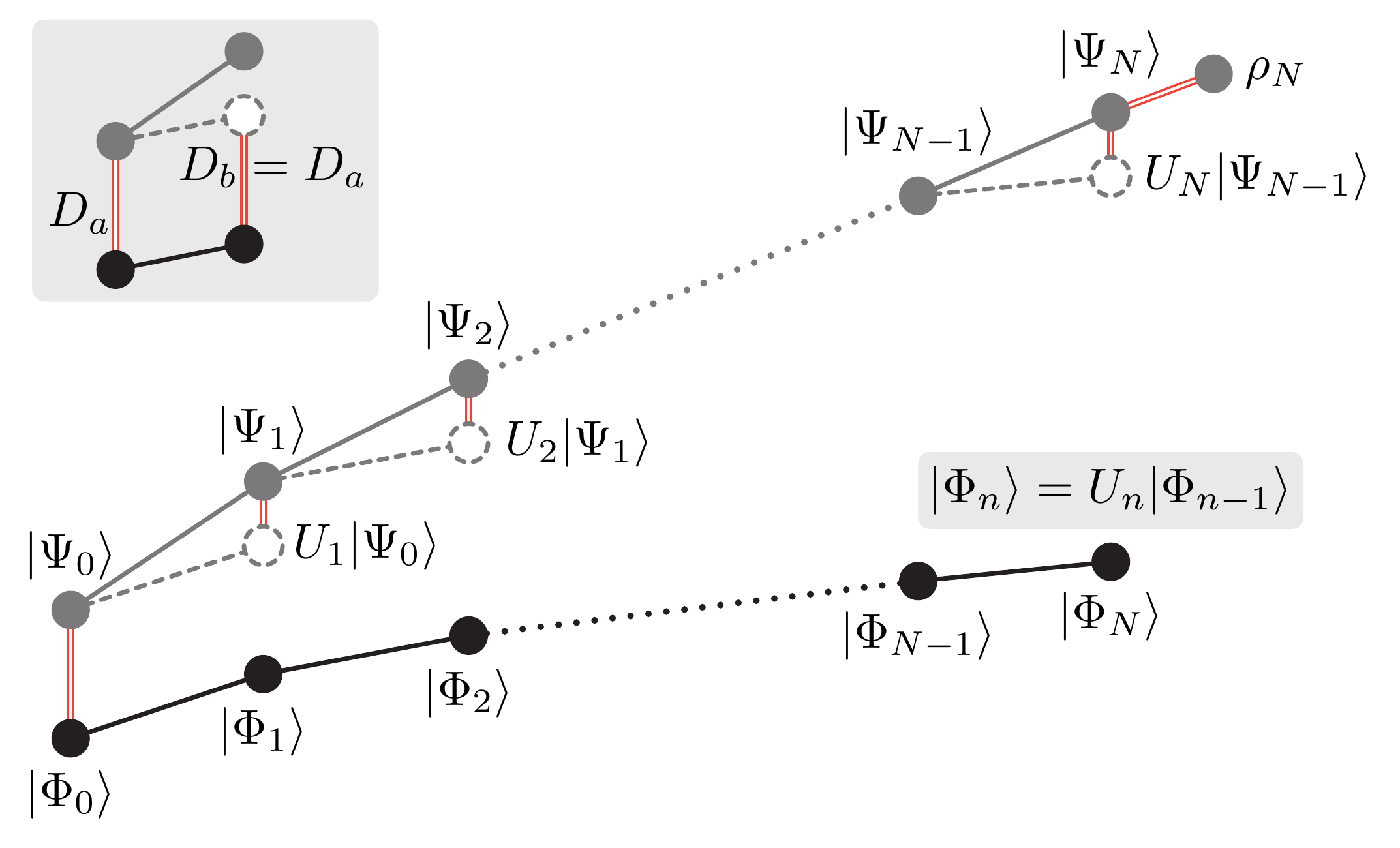}
\caption{
The distance between the true wavefunction and the wavefunction obtained from the hybrid algorithm. Black circles denote true wavefunctions given by the exact time evolution at the time $t_0,t_1,\ldots,t_N$, respectively. $U_n$ denotes the exact time evolution during the time from $t_{n-1}$ to $t_n$. Gray circles denote trial wavefunctions, and each empty circle with a dashed edge is the wavefunction at the time $t_{n}$ given by the exact time evolution $U_n$ and taking the trial wavefunction at the time $t_{n-1}$ as the initial state. Distances are marked by red double lines. Note that $D_a = D(\ket{\Phi_{n-1}},\ket{\Psi_{n-1}})$ and $D_b = D(\ket{\Phi_{n}},U_n\ket{\Psi_{n-1}})$. The distance between $\ket{\Phi_N}$ and $\rho_N$ is not larger than the sum of all of the red double lines in the main figure.
}
\label{fig:distance}
\end{figure}

There are four types of errors that can result in infidelity in the variational quantum simulation: i) errors due to limited generality of the trial wavefunction, which may only be able to describe the simulated system approximately; ii) errors in the numerical integration obtained by solving Eq.~(\ref{eq:DifEq}), which is always approximate because of the finite discretisation of time; iii) shot noise in measuring equation coefficients $M$ and $V$; and iv) errors due to noise in the quantum machine, e.g.~decoherence and quantum gate infidelity.

A good trial wavefunction allows us to not only reduce trial-wavefunction errors but also minimise the difficulty of implementing the algorithm, e.g.~only use small-size quantum circuits. Whether a good trial wavefunction can be found depends on the simulated system and our understanding of the physics in that system. However, a trial wavefunction that contains a polynomial number of parameters and is a high-fidelity approximation to the true wavefunction always exists. For example, one can set the trial wavefunction in the Trotter-Suzuki form, i.e.~take the sequence of gate operations $R$ in the form of Eq.~(\ref{eq:decomposition}), and then take the evolution time of each term as a variational parameter ($\{\lambda_k\} = \{\tau_{n,j}\}$) rather than a fixed value as in the Trotterisation algorithm~\cite{Wecker2015}. Using the Trotter-Suzuki-form trial wavefunction, we know that the probability of trial-wavefunction errors decreases with the number of Trotterisation slices $N_{\rm t}$ as `${\rm Error} \propto 1/N_{\rm t}$'~\cite{Dhand2014} in the worst case.

Integration errors depend on the numerical method for solving the differential equation~(\ref{eq:DifEq}). We take the Euler method as an example. In the Euler method, the probability of error is proportional to the size of each step $\delta t$~\cite{Butcher2008}. Therefore, by choosing a small step size, integration errors can be suppressed.

Shot noise and machine noise occur in the implementation of the algorithm while trial-wavefunction errors and integration errors are due to the imperfection of the algorithm itself. The effect of implementation errors in the integration process is that coefficients $M$ and $V$ evaluated using the quantum computer are inaccurate, i.e.~their values are different from their true values $M^{(0)}$ and $V^{(0)}$ given by Eqs.~(\ref{eq:M})~and~(\ref{eq:V}) (that is, the values given by a quantum computer without shot noise and machine noise). If coefficients $M$ and $V$ are inaccurate, time derivatives $\{\dot{\lambda}_{k}\}$ obtained from $M$ and $V$ [see Eq.~(\ref{eq:DifEq})] are different from their true values $\{\dot{\lambda}_{k}^{(0)}\}$ obtained from $M^{(0)}$ and $V^{(0)}$. Using $\delta M = M - M^{(0)}$, $\delta V = V - V^{(0)}$ and $\delta \dot{\lambda} = \dot{\lambda} - \dot{\lambda}^{(0)}$ to denote deviations from true values and using $\dot{\lambda} = g(M,V)$ to denote the solution of Eq.~(\ref{eq:DifEq}), we have $\delta \dot{\lambda}_k \simeq \sum_{p,q}\frac{\partial g_k}{\partial M_{p,q}}\delta M_{p,q} + \sum_{q}\frac{\partial g_k}{\partial V_q}\delta V_q$. When the matrix is invertible in the vicinity of $M^{(0)}$, $\delta \dot{\lambda} \simeq - \frac{1}{M^{(0)}} \delta M \dot{\lambda}^{(0)} + \frac{1}{M^{(0)}} \delta V$.

\subsection{Trace distance}

To analyse errors in the hybrid algorithm, we use the trace distance $D(\rho,\rho') = \frac{1}{2}\Tr \abs{\rho - \rho'}$~\cite{Nielsen2010} as the measure of error severity. The degree of error in the overall process is given by $D(\ket{\Phi(t_N)},\rho_N)$. Here, $\ket{\Phi_n} \equiv \ket{\Phi(t_n)}$ denotes the true wavefunction, $\ket{\Psi_n} \equiv \ket{\Psi(t_n)}$ denotes the trial wavefunction, and $\rho_N$ is the state prepared in the quantum computer according to the state $\ket{\Psi_N}$. The two states $\rho_N$ and $\ket{\Psi_N}$ are different because of the machine noise. The triangle inequality holds for the trace distance, i.e.~$D(\rho,\rho') \leq D(\rho,\rho'') + D(\rho'',\rho')$. Therefore, an upper bound of $D(\ket{\Phi_N},\rho_N)$ is given by (see Fig.~\ref{fig:distance})
\begin{eqnarray}
D(\ket{\Phi_N},\rho_N) &\leq & D(\ket{\Phi_0},\ket{\Psi_0}) + D(\ket{\Psi_N},\rho_N) \notag \\
&& + \sum_{n=1}^N D(U_n\ket{\Psi_{n-1}},\ket{\Psi_n}).
\label{eq:bound1}
\end{eqnarray}
Here, $U_n$ is the exact evolution during the time from $t_{n-1}$ to $t_n$ ($U_n = e^{-iH\delta t}$ if the Hamiltonian is time-independent), therefore $\ket{\Phi_n} = U_n \ket{\Phi_{n-1}}$. To obtain the upper bound, we have used $D(\ket{\Phi_{n-1}},\ket{\Psi_{n-1}}) = D(U_n\ket{\Phi_{n-1}},U_n\ket{\Psi_{n-1}}) = D(\ket{\Phi_{n}},U_n\ket{\Psi_{n-1}})$.

To distinguish algorithm errors and implementation errors, we use the inequality $D(U_n\ket{\Psi_{n-1}},\ket{\Psi_n}) \leq D(U_n\ket{\Psi_{n-1}},\ket{\Psi^{(0)}_n}) + D(\ket{\Psi^{(0)}_n},\ket{\Psi_n})$. Here, the state $\ket{\Psi^{(0)}_n}$ is a trial state corresponding to parameters $\{\lambda_{k}^{(0)} = \lambda_{k}(t_{n-1}) + \dot{\lambda}_{k}^{(0)} \delta t\}$, i.e.~assuming that the quantum computer reports accurate values of $M$ and $V$ at the time $t_{n-1}$. Then, the upper bound can be rewritten as
\begin{eqnarray}
D(\ket{\Phi_N},\rho_N) \leq D_{\rm A} + D_{\rm I},
\label{eq:bound2}
\end{eqnarray}
where
\begin{eqnarray}
D_{\rm A} &=& D(\ket{\Phi_0},\ket{\Psi_0}) + \sum_{n=1}^N D(U_n\ket{\Psi_{n-1}},\ket{\Psi^{(0)}_n}), \\
D_{\rm I} &=& \sum_{n=1}^N D(\ket{\Psi^{(0)}_n},\ket{\Psi_n}) + D(\ket{\Psi_N},\rho_N).
\end{eqnarray}
Here, $D_{\rm A}$ corresponds to algorithm errors, and $D_{\rm I}$ corresponds to implementation errors.

Algorithm errors in each time step can be expressed as
\begin{eqnarray}
&& D(U_n\ket{\Psi_{n-1}},\ket{\Psi^{(0)}_n}) = \sqrt{\Delta^{(2)}\delta t^2 + E},
\label{eq:AE}
\end{eqnarray}
where
\begin{eqnarray}
\Delta^{(2)} &=& \braket{\delta \Psi_{n}}{\delta \Psi_{n}} - \abs{ \braket{\delta \Psi_{n}}{\Psi_{n-1}}}^2,
\label{eq:D2}
\end{eqnarray}
\begin{eqnarray}
\ket{\delta \Psi_{n}} &=& -iH\ket{\Psi_{n-1}} - \sum_k \dot{\lambda}_{k}^{(0)} \frac{\partial \ket{\Psi_{n-1}}}{\partial \lambda_{k}},
\end{eqnarray}
and
\begin{eqnarray}
\abs{E} &\leq & \left( \sum_{m = 0}^\infty \frac{\norm{H^m} \delta t^m}{m!} \right)^2
\left( \sum_{m = 0}^\infty \frac{\norm{\frac{d^m R}{dt^m}} \delta t^m}{m!} \right)^2 \notag \\
&&- \left(1 + \norm{H}\delta t + \norm{\frac{dR}{dt}}\delta t \right)^2 \notag \\
&&- \left( \norm{H^2} + 2\norm{H}\norm{\frac{dR}{dt}} + \norm{\frac{d^2 R}{dt^2}} \right)\delta t^2 \notag \\
&=& \Delta^{(3)} \delta t^3 + \mathcal{O}(\delta t^4).
\label{eq:E}
\end{eqnarray}
Here, we have used Taylor expansions of $U_n\ket{\Psi_{n-1}}$ and $\ket{\Psi^{(0)}_n}$, i.e.~$U_n\ket{\Psi_{n-1}} = \sum_{m=0}^\infty (\delta t^m / m!) (-iH)^m \ket{\Psi_{n-1}}$ and $\ket{\Psi^{(0)}_n} = \sum_{m=0}^\infty (\delta t^m / m!) (d^m R/dt^m) \ket{\bar{0}}$, where $\frac{d}{dt} = \sum_k \dot{\lambda}_{k}^{(0)} \frac{\partial}{\partial \lambda_{k}}$. See Appendix~\ref{sec:AlgErr} for details. The matrix norm is induced by the vector norm, therefore $\abs{ \bra{\Psi_{n-1}} H^{m'} \frac{d^m R}{dt^m} \ket{\bar{0}} } \leq \norm{H^{m'}} \norm{\frac{d^m R}{dt^m}}$.

Implementation errors in each time step are due to the difference between time derivatives obtained from the real quantum computer and their true values, i.e.~$\delta \dot{\lambda}_{k} = \dot{\lambda}_{k} - \dot{\lambda}^{(0)}_{k}$, which can be expressed as
\begin{eqnarray}
&& D(\ket{\Psi^{(0)}_n},\ket{\Psi_n}) = \sqrt{ \delta \dot{\lambda}^{\rm T} A \delta \dot{\lambda}\delta t^2 + \mathcal{O}(\delta t^3)},
\end{eqnarray}
where $A$ is a positive semi-definite matrix and
\begin{eqnarray}
A_{k,q} &=& \frac{\partial \bra{\Psi_{n-1}}}{\partial \lambda_q}\frac{\partial \ket{\Psi_{n-1}}}{\partial \lambda_k} \notag \\
&&- \frac{\partial \bra{\Psi_{n-1}}}{\partial \lambda_q}\ketbra{\Psi_{n-1}}{\Psi_{n-1}}\frac{\partial \ket{\Psi_{n-1}}}{\partial \lambda_k}
\end{eqnarray}
Here, we have used the Taylor expansion of $\ket{\Psi_n}$, which is the same as $\ket{\Psi^{(0)}_n}$ but with $\{\dot{\lambda}_{k}^{(0)}\}$ replaced by $\{\dot{\lambda}_{k}\}$.

\subsection{Cost of the hybrid algorithm}

Using $Q_{\rm max}$ to denote the maximum value of the quantity $Q$ for all $t_n$, we have
\begin{eqnarray}
D_{\rm A} &\lesssim & D(\ket{\Phi_0},\ket{\Psi_0}) + \sqrt{\Delta^{(2)}_{\rm max}}T + \sqrt{\Delta^{(3)}_{\rm max} \delta t}T \\
D_{\rm I} &\lesssim & \sqrt{\norm{A}_{\rm max}}\norm{\delta \dot{\lambda}}_{\rm max}T + D(\ket{\Psi_N},\rho_N).
\end{eqnarray}
The first term of $D_{\rm A}$ and the second term of $D_{\rm I}$ are due to imperfections in approximating the initial state using the trial wavefunction and preparing the final state in the quantum computer with machine noise, respectively. Note that these two terms are not $T$ dependent; for simulations over a substantial time $T$ we may expect them to make relatively small contributions. In the following, we analyse the other three terms one by one.

The second term of $D_{\rm A}$ is the accumulation of trial-wavefunction imperfections in each time step. Using the circuit in Fig.~\ref{fig:circuit}, every term in $\Delta^{(2)}$ can be measured by a method analogous to that used for obtaining $M$ and $V$. The accuracy is again limited by the shot noise and machine noise. Therefore, algorithm errors due to the trial wavefunction can be continually estimated during the execution of the hybrid quantum computation.

The third term of $D_{\rm A}$ is caused by the finite integration step size, which can be reduced by decreasing $\delta t$. In order to limit this term to $\varepsilon$, we need to choose a step size $\delta t \sim \varepsilon^2/(\Delta^{(3)}T^2)$, i.e.~the number of time steps $N \sim \Delta^{(3)} T^3 / \varepsilon^2$. When the trial wavefunction is a good approximation to the exact state, we can expect that $\frac{d^m R}{dt^m}\ket{\Psi} \simeq H^{m}\ket{\Psi}$, which implies $\norm{\frac{d^m R}{dt^m}} \sim \norm{H^{m}}$ in the subspace of the problem, and in this case $\Delta^{(3)} \sim \norm{H}^{3}$.

The first term of $D_{\rm I}$ is due to the difference between $\{\dot{\lambda}_{k}\}$ and their true values $\{\dot{\lambda}_{k}^{(0)}\}$. The difference is $\norm{\delta \dot{\lambda}} \leq \norm{\frac{1}{M^{(0)}}}^2 \norm{V^{(0)}} \norm{\delta M} + \norm{\frac{1}{M^{(0)}}} \norm{\delta V}$. Here, $M$ and $V$ are evaluated as required by the algorithm. Similar to $\Delta^{(2)}$, each element of $A$ can also be measured using the circuit in Fig.~\ref{fig:circuit}. Therefore, the susceptibility to shot noise and machine noise in the integration process can be estimated during the execution of the hybrid quantum computation. The algorithm is susceptible to implementation errors when $M^{(0)}$ is singular, which should be avoided when choosing the trial wavefunction. As a worst-case scenario, implementation errors accumulate linearly with the simulated time. However, it can be far less severe: In Sec.~\ref{sec:numerics}, we will explore an example in which the accumulation of errors due to the machine noise is almost negligible compared with errors in a quantum simulation based on the conventional Trotter-Suzuki decomposition.

Shot noise can be suppressed by repeating quantum circuits for measuring $M$ and $V$ many times. To measure the quantity $\mean{X}$, the deviation due to the shot noise decreases with the number of repetitions $N_{\rm r}$ as $\delta \mean{X} \propto 1/\sqrt{N_{\rm r}}$. Therefore, if there is only shot noise, we have $\norm{\delta \dot{\lambda}} \sim \Delta /\sqrt{N_{\rm r}}$, where $\Delta = \norm{\frac{1}{M^{(0)}}}^2 \norm{V^{(0)}} \Theta_M + \norm{\frac{1}{M^{(0)}}} \Theta_V$, $\Theta_M = 2[\sum_{k,q}(\sum_{i,j}\abs{i f^*_{k,i}f_{q,j}})^2]^{1/2}$ and $\Theta_V = 2[\sum_{k}(\sum_{i,j}\abs{f^*_{k,i}h_{j}})^2]^{1/2}$. In order to limit the overall effect of shot noise to $\varepsilon'$, we need to choose $N_{\rm r} \sim \norm{A}\Delta^2 T^2 / \varepsilon^{\prime 2}$. 

The number of distinct circuits $N_{\rm c}$ required for finding the $M$ and $V$ parameters depends on $N_{\rm H}$ the number of terms in the Hamiltonian $H$ [see Eq.~(\ref{eq:Ham})], $N_{\rm d}$ the number of terms in each time derivative of $R_k$ [see Eq.~(\ref{eq:dRk})], and $N_{\rm v}$ the number of parameters in the trial wavefunction. Note that $N_{\rm c} = N_{\rm v}^2 N_{\rm d}^2 + N_{\rm v} N_{\rm d} N_{\rm H}$, where the first term corresponds to $M$, and the second term corresponds to $V$. If each $R_k$ is realised by $N_R$ gates, and each term $\sigma_i$ or $\sigma_{k,i}$ in the Hamiltonian or the time derivative of $R_k$ is a Pauli operator of less than $K$ qubits, each circuit includes at most $N_{\rm g} = N_{\rm v}N_R + 2(K+1)$ gates. Here, we have assumed that each controlled-$U$ gate in the circuit (Fig.~\ref{fig:circuit}) is realised by $K$ two-qubit controlled-$\sigma$ gates, where $\sigma$ is a single-qubit Pauli operator. To complete the circuit, other operations include preparing initial states of the ancillary qubit and the register and measuring the ancillary qubit.

The overall computation includes $N$ times steps, in each time step, $N_{\rm c}$ circuits are implemented, each circuit contains $N_{\rm g}$ gates and is repeated $N_{\rm r}$ times. Therefore, the overall number of gates is
\begin{eqnarray}
N N_{\rm c} N_{\rm g} N_{\rm r}
&\sim & \frac{\norm{A} \Delta ^2 \Delta^{(3)} T^5}{\varepsilon^2 \varepsilon^{\prime 2}} \notag \\
&&\times N_{\rm v} N_{\rm d}(N_{\rm v} N_{\rm d} +  N_{\rm H}) \notag \\
&&\times [N_{\rm v}N_R + 2(K+1)].
\end{eqnarray}
From this expression, we see that the cost is a polynomial function with respect to the integration error $\varepsilon$, the shot-noise error $\varepsilon'$ and the simulated time $T$. Factors $\norm{A}$, $\Delta$ and $\Delta^{(3)}$ depend on the form of the trial wavefunction. However, during the actual execution of the hybrid computation, it will be possible to estimate $\norm{A}$ and $\Delta$. Moreover $\Delta^{(3)} \sim \norm{H}^{3}$ when the trial wavefunction is a good approximation to the true state.

It is important to remember that while the overall gate count will be a large (albeit polynomially-scaling) total, the complete calculation is formed of many small quantum calculations of depth $N_{\rm g}$. Each small computation is isolated from the others, i.e. there is no shared or persistent quantum resource, and indeed $\sim N_{\rm c} N_{\rm r}$ such circuits could be performed in parallel using that many separate small quantum computers.

\section{Effect of machine noise and error reduction}
\label{sec:noise}

\begin{figure}[tbp]
\centering
\includegraphics[width=1\linewidth]{\figpath 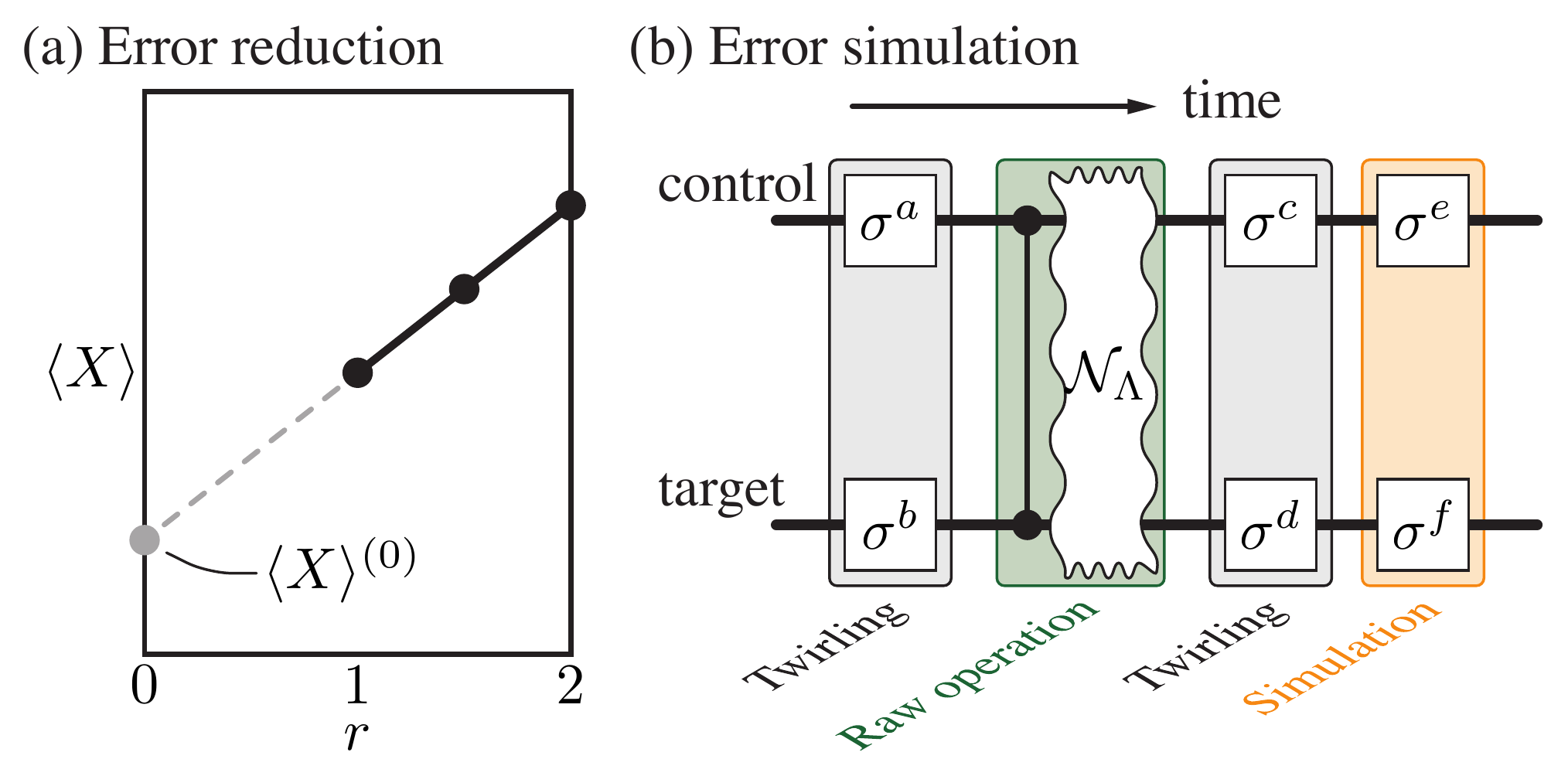}
\caption{
(a) A schematic diagram of the error reduction. The true value $\mean{X}^{(0)}$ is inferred by measuring $\mean{X}$ for a set of error factors $r$ and fitting data using the function $\mean{X} = \mean{X}^{(0)} + \chi r$. (b) Error twirling and simulation. Non-stochastic errors in a controlled-phase gate can be converted into stochastic errors by performing Pauli gates before and after the gate, and error probabilities can be tuned by applying Pauli gates after the gate.
}
\label{fig:error}
\end{figure}

Of the implementation errors, machine noise is the more problematic. Shot noise can be suppressed merely by repeating each quantum circuit many times, and the number of repetitions is a polynomial function with respect to the accuracy. Machine noise is less easily dismissed, and it is the focus of this section.

Machine noise need not necessarily result in computing errors. The task of the quantum computer is to evaluate coefficients $M$ and $V$, so that the classical computer can solve time derivatives $\{\dot{\lambda}_{k}\}$ according to Eq.~(\ref{eq:DifEq}). Machine noise (as well as shot noise) can potentially cause computing errors by changing these time derivatives from their true values $\{\dot{\lambda}_{k}^{(0)}\}$. However, sometimes machine noise does not change values of $\{\dot{\lambda}_{k}\}$. For example, consider the case in which the effect of machine noise is to depolarise the ancillary qubit at a fixed level, i.e.~the output of the quantum computer becomes $\mean{X} = \eta \mean{X}^{(0)}$, where $\eta$ is a constant independent of the quantum circuit. In this case, all equation coefficients are scaled as $M = \eta M^{(0)}$ and $V = \eta V^{(0)}$. As long as $\eta$ is non-zero, the solution $\{\dot{\lambda}_{k}\}$ of Eq.~(\ref{eq:DifEq}) is the same for any value of $\eta$. Therefore, only an inhomogeneous scaling of quantum outputs $\mean{X}$ results in computing errors.

An example of the homogeneous scaling is the case of balanced measurement errors. Errors in the measurement on the ancillary qubit (see Fig.~\ref{fig:circuit}) can be modelled as follows: if the state of the qubit is $\ket{0}$ ($\ket{1}$), the measurement outcome is correct, i.e.~$0$ ($1$), with the probability $1-p_{0}$ ($1-p_{1}$), and the outcome is incorrect, i.e.~$1$ ($0$), with the probability $p_{0}$ ($p_{1}$). If there are no other implementation errors, quantum outputs are changed from $\mean{X}^{(0)}$ to $\mean{X} = (p_1-p_0) + (1-p_0-p_1)\mean{X}^{(0)}$ (the measurement in the $\{\ket{+},\ket{-}\}$ basis is done by performing a Hadamard gate before measuring the qubit in the $\{\ket{0},\ket{1}\}$ basis). If measurement errors are balanced, i.e.~$p_{0} = p_{1}$, the effect of measurement errors is a fixed scaling factor $\eta = 1-p_0-p_1$, which does not result in computing errors. Therefore, our hybrid algorithm is inherently insensitive to measurement errors on the ancillary qubit if these errors are balanced. We remark that if single-qubit gates are reliable, one can flip the qubit before the measurement, so that measurement errors are effectively balanced.

Measurement errors can be corrected even if they are not balanced. If $p_0$ and $p_1$ can be evaluated by benchmarking measurement operations, one can easily work out the true value $\mean{X}^{(0)}$ using the value obtained from the real machine: $\mean{X}^{(0)} = [\mean{X} - (p_1-p_0)] / (1-p_0-p_1)$. We remark that when error probabilities are higher, the denominator is smaller, which means that we need to evaluate $\mean{X}$ with a higher accuracy in order to achieve the same accuracy of $\mean{X}^{(0)}$. Next, we show that a similar procedure can be applied to any machine noise if errors due to the machine noise are stochastic with tunable probabilities. We also show how to convert errors in two-qubit entangling gates, which are expected to be the main sources of errors, into stochastic errors if they are not stochastic, and how to simulate stochastic errors to tune error probabilities.

\subsection{Error reduction}

\begin{figure*}[tbp]
\centering
\includegraphics[width=1\linewidth]{\figpath 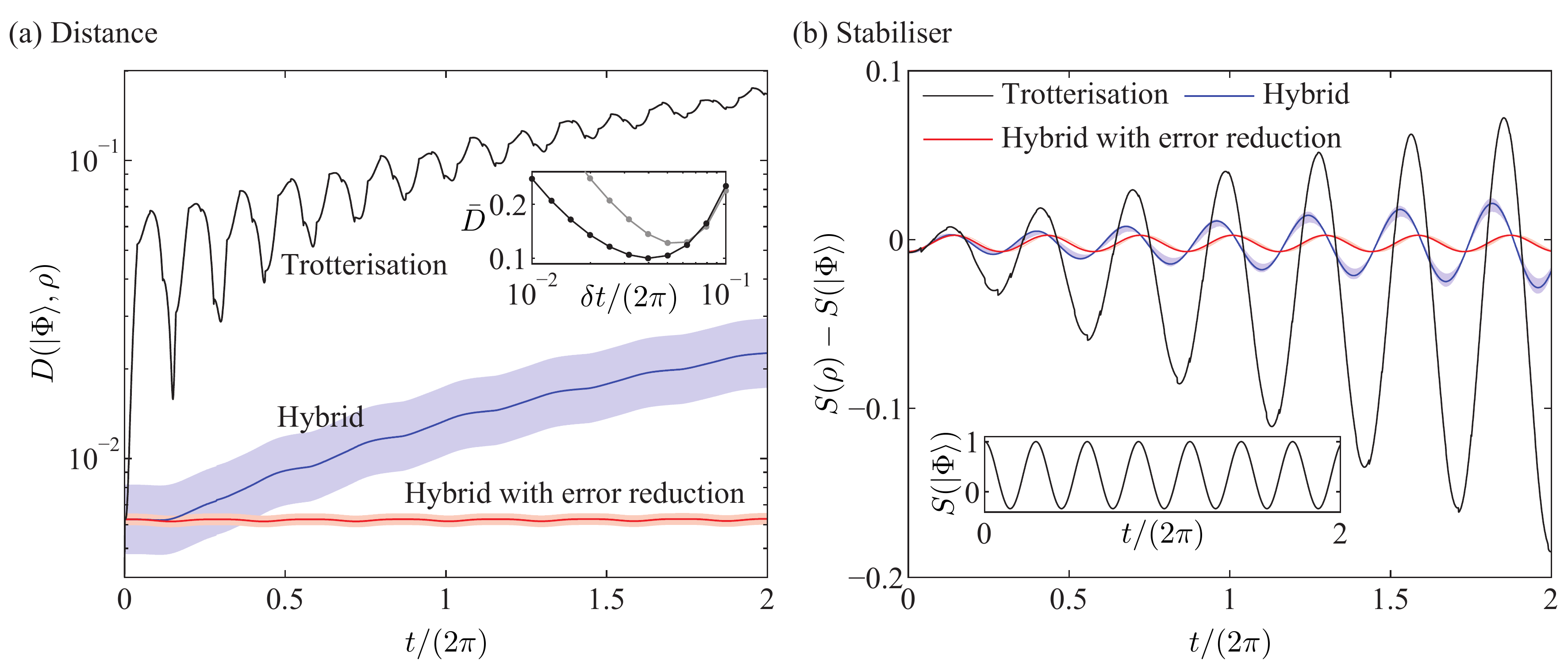}
\caption{
Numerical estimations of the performance of quantum simulation using the Trotterisation algorithm and the hybrid algorithm. The simulated system is a quantum Ising model of three spins initialised in the cluster state. (a) Trace distance $D(\ket{\Phi},\rho)$ as a function of the simulated time $t$. Here, $\ket{\Phi(t)}$ is the true state and $\rho(t)$ is the state prepared in the quantum computer. The Trotterisation algorithm (black curves) and the hybrid algorithm are compared. In the Trotterisation algorithm $\rho(t)$ is prepared according to the Trotter-Suzuki decomposition, and in the hybrid algorithm $\rho(t)$ is prepared according to the trial state $\ket{\Psi(t)}$. We have taken the error rate $\epsilon^{(2)} = 0.1\%$. The hybrid algorithm without the error reduction (blue curves) is already much more reliable than the conventional Trotterisation algorithm. In the hybrid algorithm, one can further reduce the distance using the error reduction protocol (red curve). The residual distance that cannot be eliminated using the error reduction is mainly due to errors in the state preparation (i.e.~$D(\vert \Psi_N \rangle,\rho_N)$). Inset: the parameter $\delta t$ in the Trotterisation algorithm is optimised to minimise the average distance, and the gray curve is obtained by using the lowest-order symmetric Trotter-Suzuki decomposition (see Appendix~\ref{sec:DetNum}). (b) The difference between the average value of a stabiliser estimated using the quantum computer and its true value. Here, $S(\rho) = \Tr(S_2\rho)$ and $S(\ket{\Phi}) = \bra{\Phi}S_2\ket{\Phi}$. Average values of other stabilisers are only slightly different. The true average value of the stabiliser is plotted in the inset. In both figures (a) and (b), light blue and red bands denote the fluctuation due to shot noise: with $68\%$ chance (in total 100 trials), the corresponding quantity is within the band.
}
\label{fig:plot}
\end{figure*}

Errors in an operation are stochastic if the operation is described by a superoperator $\mathcal{N} \mathcal{U}$ and $\mathcal{N}$ has the form $\mathcal{N} = (1-\epsilon)\mathcal{I} + \epsilon\mathcal{E}$. Here, $\mathcal{U}$ is the ideal operation without errors, $\mathcal{N}$ is the superoperator describing the effect of the noise, $\mathcal{I}$ is an identity operation, and errors $\mathcal{E}$ occur with the probability $\epsilon$. Here, $\mathcal{E}$ is a valid quantum operation, i.e.~trace-preserving completely positive map.

Given an initial state $\ket{\bar{0}}$, after a sequence of operations, the final state of the quantum computer is
\begin{eqnarray}
\rho = \mathcal{N}_{L} \mathcal{U}_{L} \cdots \mathcal{N}_{l} \mathcal{U}_{l} \cdots \mathcal{N}_{1} \mathcal{U}_{1} (\ketbra{\bar{0}}{\bar{0}}),
\end{eqnarray}
where $\mathcal{N}_{l} \mathcal{U}_{l}$ denotes the $l$th operation. Taking into account the fact that errors are stochastic, the quantum outcome can be rewritten in the form
\begin{eqnarray}
\mean{X} = (1 - r\sum_l \epsilon_l)\mean{X}^{(0)} + r\mean{X}^{(1)} + \mathcal{O}(r^2).
\label{eq:expansion}
\end{eqnarray}
Here, $\mean{X} = \Tr(X\rho)$, $\mean{X}^{(0)} = \Tr(X\rho^{(0)})$ is the value without errors,
\begin{eqnarray}
\rho^{(0)} = \mathcal{U}_{L} \cdots \mathcal{U}_{l} \cdots \mathcal{U}_{1} (\ketbra{\bar{0}}{\bar{0}}),
\end{eqnarray}
$\mean{X}^{(1)} = \Tr(X\rho^{(1)})$, and
\begin{eqnarray}
\rho^{(1)} = \sum_l \epsilon_l \mathcal{U}_{L}  \cdots \mathcal{E}_{l} \mathcal{U}_{l} \cdots  \mathcal{U}_{1} (\ketbra{\bar{0}}{\bar{0}}).
\end{eqnarray}
The $l^\text{th}$ term of $\rho^{(1)}$ corresponds to the case in which only the $l^\text{th}$ operation causes errors and all other operations are ideal. Note that in these equations we have replaced error probabilities $\epsilon_l$ with $r\epsilon_l$, where $r$ is a convenient scale factor allowing us to write  $\mean{X}(r)$ as a function of $r$, and $\mean{X}(0) = \mean{X}^{(0)}$.

If probabilities of errors are tunable, we can infer the value of $\mean{X}(0)$ by measuring values of $\mean{X}(r)$ of a set of different factors $r$ [see Fig.~\ref{fig:error}(a)]. Clearly $r$ can never be zero for our machine, as this would imply that we have the ability to fully switch off the machine noise, making it perfect. If $\{\epsilon_l\}$ are the minimum error probabilities allowed by the machine, the minimum value of $r$ achievable by the machine is $1$. To infer the value of $\mean{X}(0)$, firstly, we take $N_{\rm X}$ values of $r$ as $r_1, r_2, \ldots, r_{N_{\rm X}}$ and measure $\mean{X}(r_1), \mean{X}(r_2), \ldots, \mean{X}(r_{N_{\rm X}})$ using the quantum computer, where we can take $r_1 = 1$. For example, in Fig.~\ref{fig:error}(a), we have taken $r_1 = 1$, $r_2 = 1.5$ and $r_3 = 2$. Secondly, we fit quantum outputs using the function $\mean{X}(r) = \mean{X}^{(0)} + \chi r$, where $\chi = - \mean{X}^{(0)}\sum_l \epsilon_l+ \mean{X}^{(1)}$. As a result of the fitting, we obtain the value of $\mean{X}^{(0)}$, represented by the gray circle in Fig.~\ref{fig:error}(a). In this way, the first-order contribution of machine noise can be corrected. Similarly, by considering second-order terms in the expansion~(\ref{eq:expansion}), one can fit data using a function with second-order terms (i.e.~$r^2$ terms) to correct the second-order contribution of machine noise. Using the extrapolation, we can reduce the effect of the machine noise. However, the final estimation of $\mean{X}^{(0)}$ may still be different from its actual value, and the error in the extrapolation depends on the shot noise in estimating each $\mean{X}(r)$.

The error reduction protocol only works for small-size circuits, which are used in the hybrid algorithm while the Trotterisation algorithm usually needs large-size circuits. The true value $\mean{X}^{(0)}$ can be inferred because the contribution of high-order terms is much smaller than the contribution of lower-order terms, i.e.~$\abs{\mathcal{O}(r^2)} \ll \abs{r\mean{X}^{(1)}}$. The total rate of errors in the quantum circuit with $N_{\rm g}$ gates is $\sim 1-(1-\epsilon)^{N_{\rm g}} = N_{\rm g}\epsilon + N_{\rm g}^2\epsilon^2/2 + \cdots$. The first term in the expansion corresponds to the first order contribution $\abs{r\mean{X}^{(1)}}$, and so on. Therefore, high-order terms cannot be neglected if $N_{\rm g}\epsilon \gtrsim 1$. When there are too many gates in the circuit or the error rate is too high, the quantum state will be populated with errors and one cannot retrieve the true value $\mean{X}^{(0)}$ even if we consider high-order terms in the interpolation. The best experiments to date~\cite{Lucas, Wineland} have reduced two-qubit gate infidelity to the range $10^{-3}$ to $10^{-4}$. With hardware of that kind, our protocol could support hundreds of gates. Thus a simulation of a quantum system using a trial wavefunction with hundred(s) of parameters, may be feasible.

Implementing the error reduction protocol requires knowledge of the inherent machine noise. Therefore, quantum operations need to be benchmarked, e.g.~using quantum process tomography~\cite{Nielsen2010, Bialczak2010}, before the quantum coprocessor is used, and the nature of the machine noise should not vary significantly during the simulation. Alternatively, one can monitor the machine noise in the process by stopping the protocol and benchmarking operations, because in the hybrid algorithm the quantum computer only performs small-size circuits and can be stopped at any stage.

\subsection{Error twirling and simulation}

Non-stochastic errors can be converted into stochastic Pauli errors using some redundant Pauli gates~\cite{Knill2004, Wallman2015, OGorman2016}. It is a common feature of quantum computing systems that two-qubit gates are the main source of errors, i.e.~probabilities of errors in two-qubit gates are much higher than probabilities of errors in single-qubit gates~\cite{Barends2014, Harty2014, Lucas, Wineland, Ladd2010}. In this case, arbitrary errors in two-qubit gates can be converted into stochastic errors through the use of redundant Pauli gates, without introducing significant additional noise. We consider the controlled-phase gate $\Lambda = (\openone + \SIG{\rm z}{\rm c})/2 + \SIG{\rm z}{\rm t}(\openone - \SIG{\rm z}{\rm c})/2$ as an example, and it is similar for other two-qubit gates, e.g.~the controlled-NOT gate. Here c and t denote the control qubit and the target qubit, respectively.

The controlled-phase gate with noise is $\mathcal{N}_\Lambda \mathcal{U}_\Lambda$, where $\mathcal{U}_\Lambda \rho = \Lambda \rho \Lambda^\dag$. In general the noise may not be in stochastic-error form, but we can always express the noise operation in the Kraus form, i.e.~$\mathcal{N}_\Lambda \rho = \sum_h E_h \rho E_h^\dag$, where $\sum_h E_h^\dag E_h = \openone$, and each two-qubit matrix can be written as $E_h = \sum _{a = 0}^3 \sum _{b = 0}^3 \alpha _{h;a,b} \SIG{a}{\rm c} \SIG{b}{\rm t}$~\cite{Nielsen2010}. Here $a,b = 0,1,2,3$ correspond to Pauli operators $\openone$, $\SIG{\rm x}{}$, $\SIG{\rm y}{}$ and $\SIG{\rm z}{}$, respectively.

In order to convert errors, Pauli gates $\{\openone, \SIG{\rm x}{}, \SIG{\rm y}{}, \SIG{\rm z}{}\}$ are randomly chosen and applied on each qubit before and after the controlled-phase gate. If the gate before the controlled-phase gate is $U$, the gate after the controlled-phase gate is restricted to be $\Lambda U \Lambda^\dag$ in order to let random Pauli gates cancel each other. Here, $U$ and $\Lambda U \Lambda^\dag$ are both two-qubit Pauli gates. The circuit is shown in Fig.~\ref{fig:error}(b), in which we take $c = a+b(3-b)(3-2a)/2$ and $d = b+a(3-a)(3-2b)/2$, so that $\SIG{c}{\rm c} \SIG{d}{\rm t} = \Lambda \SIG{a}{\rm c} \SIG{b}{\rm t} \Lambda^\dag$ up to a phase factor. As a result, the overall operation is $\bar{\mathcal{N}}_\Lambda \mathcal{U}_\Lambda$, where $\bar{\mathcal{N}}_\Lambda$ is the superoperator describing the effective noise after applying random Pauli gates. The effective noise superoperator is in the stochastic form~\cite{Emerson2007} (see Appendix~\ref{sec:twirling}), i.e.
\begin{eqnarray}
\bar{\mathcal{N}}_\Lambda &=& F_\Lambda [\openone]
+ \sum _{(a,b) \neq (0,0)} \epsilon_{a,b} [\SIG{a}{\rm c} \SIG{b}{\rm t}],
\label{eq:effective}
\end{eqnarray}
where the fidelity is $F_\Lambda = \sum _h \abs{\alpha _{h;0,0}}^2$, and error probabilities are $\epsilon_{a,b} = \sum _h \abs{\alpha _{h;a,b}}^2$. Here, $[U]$ is a superoperator, and $[U]\rho = U\rho U^\dag$. We have assumed that Pauli gates are ideal. In the case where Pauli gates are not ideal but their error probabilities are much lower than the controlled-phase gate, noise of Pauli gates will be a perturbation to the effective noise of the controlled-phase gate. Error severity in the controlled-phase gate will effectively increase by four units of the single-qubit Pauli gate error rate, and the effective noise of the controlled-phase gate may not be fully stochastic if Pauli-gate errors are non-stochastic.

To tune probabilities of errors, we can randomly perform Pauli gates after the controlled-phase gate according to desired error probabilities [see Fig.~\ref{fig:error}(b)]. Assuming that we want to tune error probabilities from $\epsilon_{e,f}$ to $r\epsilon_{e,f}$, we can perform the Pauli gate $\SIG{e}{\rm c} \SIG{f}{\rm t}$ [$(e,f)\neq (0,0)$] with the probability $(r-1)\epsilon_{e,f}$. Because we are only interested in the case where $r\epsilon_{e,f} \ll 1$, overall error probabilities are approximately $\epsilon_{e,f} + (r-1)\epsilon_{e,f}$ where the first term is due to the raw controlled-phase gate (with noise), and the second term is due to simulated errors using single-qubit Pauli gates.

\section{Numerical results --- Quantum Ising model}
\label{sec:numerics}

To demonstrate the hybrid algorithm, we numerically simulate a small quantum computer using classical computers. We take the quantum Ising model~\cite{Sachdev2011} as an example. The Hamiltonian of the model reads $H = H_{\rm Z} + H_{\rm X}$, where $H_{\rm Z} = -J \sum_{j=1}^{n_{\rm s}} \SIG{\rm z}{j}\SIG{\rm z}{j+1}$ and $H_{\rm X} = -B \sum_{j=1}^{n_{\rm s}} \SIG{\rm x}{j}$. Here, $n_{\rm s}$ is the number of spins, and $\SIG{\rm z}{n_{\rm s}+1} = \SIG{\rm z}{1}$. In our numerical simulations, we take $J=B=1/2$ and $n_{\rm s} = 3$, therefore we need at least four qubits in the quantum computer to implement the hybrid algorithm. The trial state is chosen to be
$$\ket{\Psi} = e^{i\lambda_2 H_{\rm X}} e^{i\lambda_1 H_{\rm Z}} \ket{\Phi(0)},$$
where the initial state $\ket{\Phi(0)}$ is a one-dimensional cluster state. In the cluster state, qubits are in the eigenstate of stabilisers $S_j = \SIG{\rm z}{j-1}\SIG{\rm x}{j}\SIG{\rm z}{j+1}$ ($j = 1,2,\ldots,n_{\rm s}$) with the same eigenvalue $+1$, which is prepared by performing the controlled-phase gate on each pair of nearest neighbouring qubits initialised in the state $\ket{+}$~\cite{Raussendorf2003}. The evolution of the true state is $\ket{\Phi(t)} = e^{-iHt} \ket{\Phi(0)}$. In this example, the trial state is capable of exactly matching the true state given the correct values of the parameters.

We consider a quantum computer that can initialise qubits in the state $\ket{0}$, measure a qubit in the $\{\ket{0},\ket{1}\}$ basis, and perform single-qubit and two-qubit quantum gates. Quantum gates include the Hadamard gate, Pauli gates, phase gates $e^{i\SIG{\rm z}{}\theta}$, flip gates $e^{i\SIG{\rm x}{}\theta}$, two-qubit gates $e^{i\SIG{\rm z}{1}\SIG{\rm z}{2}\theta}$, the controlled-phase gate and the controlled-NOT gate. If we have one of the three types of two-qubit gates, the other two can be efficiently realised, e.g.~the gate $e^{i\SIG{\rm z}{1}\SIG{\rm z}{2}\theta} = H_1 \Lambda e^{i\SIG{\rm x}{1}\theta} \Lambda H_1$ can be realised using two controlled-phase gates and three single-qubit gates. Here, $H_1$ is the Hadamard gate on qubit-1. We assume that all three types of two-qubit gates can be directly implemented for simplification. The state $\ket{+}$ is prepared by initialising the qubit in the state $\ket{0}$ and performing a Hadamard gate; the measurement in the $\{\ket{+},\ket{-}\}$ basis is done by performing a Hadamard gate before measuring the qubit in the $\{\ket{0},\ket{1}\}$ basis.

We model the machine noise in the quantum computer as depolarising errors. A qubit may be initialised in the incorrect state ($\ket{1}$) with the probability $\epsilon^{\rm I}$. The measurement outcome is incorrect with the probability $p_0 = p_1 = \epsilon^{\rm M}$. For single-qubit gates, the noise superoperator is
$$\mathcal{N}^{(1)} = (1-\frac{4}{3}\epsilon^{(1)})[\openone]
+ \frac{\epsilon^{(1)}}{3}\sum_{a=0}^{3}[\SIG{(a)}{}].$$
For two-qubit gates, the noise superoperator is
$$\mathcal{N}^{(2)} = (1-\frac{16}{15}\epsilon^{(2)})[\openone]
+ \frac{\epsilon^{(2)}}{15}\sum_{a=0}^{3}\sum_{b=0}^{3}[\SIG{(a)}{1}\SIG{(b)}{2}].$$
Here, $\epsilon^{(1)}$ and $\epsilon^{(2)}$ are rates of errors per gate. We assume that error rates of all single-qubit gates are the same, error rates of all two-qubit gates are the same, and error rates of single-qubit operations are only one tenth of the error rates of two-qubit gates, i.e.~$\epsilon^{\rm I} = \epsilon^{\rm M} = \epsilon^{(1)} = \epsilon^{(2)}/10$. Because the size of quantum circuits for implementing the hybrid algorithm is small, we neglect memory errors. In this model of the machine noise, errors in quantum operations are all stochastic, and error rates can be tuned by simulating errors using single-qubit Pauli gates.

Numerical simulations are performed to find the trace distance $D(\ket{\Phi},\rho)$ between the true state $\ket{\Phi(t)}$ and the state $\rho(t)$ prepared in the quantum computer according to the trial state $\ket{\Psi(t)}$ [see Fig.~\ref{fig:plot}(a)]. Because of the machine noise, $\rho(t)$ is different from $\ket{\Psi(t)}$. The average values of stabilisers can be used to describe the quality of a cluster state. The performance of quantum algorithms in estimating the average values of stabilisers is shown in Fig.~\ref{fig:plot}(b). In our numerical simulations, we have taken $\epsilon^{(2)} = 0.1\%$, which is the state-of-the-art error rate~\cite{Lucas, Wineland}. See Appendix~\ref{sec:DetNum} for some details about our numerical simulations.

We observe that in this simulation the hybrid algorithm proves to be far more reliable than the Trotterisation algorithm. The distance in the hybrid algorithm is about ten times lower than the distance in the Trotterisation algorithm, and moreover the increase of the distance as a function of time can be largely suppressed in the hybrid algorithm by using the error reduction scheme [Fig.~\ref{fig:plot}(a)]. As a result, the hybrid algorithm can provide a much better estimation of the average values of stabilisers [Fig.~\ref{fig:plot}(b)]. We would like to stress that in order to make the comparison fair, the time interval selected for the Trotterisation algorithm has been optimised to minimise errors (in practice this would be possible only if the distance from the true state can be measured in the quantum computer, which would probably negate the need for simulation). We have also considered the lowest-order symmetric Trotter-Suzuki decomposition~\cite{Berry2007, Wiebe2010}, which can reduce errors due to the Trotterisation compared with the conventional Trotterisation scheme. However, we find that the total errors are more significant using the symmetric decomposition given the gate error rate $\epsilon^{(2)} = 0.1\%$. In the hybrid algorithm, we have used the fourth-order Runge-Kutta method and taken $\delta t = 2\pi \times 10^{-6}$ to eliminate errors due to the numerical integration. We have neglected the effect of shot noise, therefore all errors are due to machine noise.

Given a finite time cost of implementing the hybrid algorithm, we need to consider the effect of errors in the numerical integration and shot noise. Taking into account the effect of these two types of imperfections, we find that the distance in the hybrid algorithm may be increased but is still much lower than the distance in the Trotterisation algorithm [blue and red bands in Fig.~\ref{fig:plot}(a)]: we take $\delta t = 2\pi \times 10^{-4}$ and assume that each circuit is repeated for $N_{\rm r} = 10^4$ ($N_{\rm r} = 10^6$) times to measure $\mean{X}$ in the hybrid algorithm (with the error reduction). Increasing the distance only slightly changes the estimation of average values of stabilisers [Fig.~\ref{fig:plot}(b)].

\section{Summary}
\label{sec:summary}

We have proposed a quantum algorithm for simulating the time evolution of a quantum system. In this algorithm, both a classical processor and a quantum coprocessor are tightly integrated. Because of the assistance of the classical computer, the algorithm can be implemented with quantum circuits of much less depth (i.e. fewer quantum operations) compared with the canonical Trotterisation algorithm. We discussed the robustness of the algorithm to noise, and we found that the algorithm can automatically correct some errors induced by the noise in the quantum computer. Moreover, by deliberately amplifying stochastic noise in a controllable way, the zero-error limit can be estimated; consequently, the effect of errors can be significantly suppressed without the need for code-based quantum error correction and its concomitant resource overheads. This quantum algorithm can also be parallelised easily. The task of the quantum coprocessor is to repeatedly implement a set of small-size quantum circuits, therefore the computing speed can be accelerated by using a cluster of quantum coprocessors, in which each works independently and there are no quantum channels linking them. 

In view of these various merits, we believe that our algorithm is a promising candidate for early-stage non-fault-tolerant quantum computers.

\begin{acknowledgments}
This work was supported by the EPSRC National Quantum Technology Hub in Networked Quantum Information Technologies. The authors would like to acknowledge the use of the University of Oxford Advanced Research Computing (ARC) facility in carrying out this work. http://dx.doi.org/10.5281/zenodo.22558.
\end{acknowledgments}

\bigskip

{\bf \noindent Note added:} \newline Recently, a highly relevant paper was also posted by Temme, Bravyi, and Gambetta (arXiv:1612.02058). In that work the authors also studied an error mitigation strategy involving deliberate variation of the error severity and subsequent extrapolation to the most likely zero-error value of their observable. While the authors analyse this as a general technique and investigate higher order corrections, rather than employing the technique in the specific context of quantum dynamical simulation as we do here, the results of our two papers are consistent and can be compared.

\appendix

\section{Simulated model of Fig. 1(b)}
\label{sec:model}

In Fig.~\ref{fig:hybrid}(b), the time evolution of a qubit is simulated. The time evolution is driven by the Hamiltonian $H = -(\sigma^{\rm y} + \sigma^{\rm z} \cos t - \sigma^{\rm y} \sin t)/2$. The trial state is in the form $\ket{\Psi} = e^{i(\pi/2) \lambda_2 \sigma^{\rm z}} e^{i(\pi/2) \lambda_1 \sigma^{\rm y}} \ket{0}$. The initial state is given by $\lambda_1 = 3/4$ and $\lambda_2 = -1/2$, and $T = 2\pi$. To Demonstrate this example, we need a quantum computer of two qubits.

\section{Algorithm errors}
\label{sec:AlgErr}

Algorithm errors in each time step are expressed as per Eq.~(\ref{eq:AE}), where $E = 1 - \abs{\bra{\Psi^{(0)}_n}U_n\ket{\Psi_{n-1}}}^2 - \Delta^{(2)}\delta t^2$. Using Taylor expansions of $\ket{\Psi^{(0)}_n}$ and $U_n\ket{\Psi_{n-1}}$ and the expression of $\Delta^{(2)}$ in Eq.~(\ref{eq:D2}), we find that
\begin{eqnarray}
\abs{\bra{\Psi^{(0)}_n}U_n\ket{\Psi_{n-1}}}^2 = 1 - \Delta^{(2)}\delta t^2 + \mathcal{O}(\delta t^3).
\end{eqnarray}
Here we used the observation that $\Re \bra{\bar{0}} \frac{dR^\dag}{dt} \ket{\Psi_{n-1}} = 0$ and $\Re \bra{\bar{0}} \frac{d^2R^\dag}{dt^2} \ket{\Psi_{n-1}} + \norm{\frac{dR}{dt} \ket{\bar{0}}}^2 = 0$. We remark that $\ket{\Psi_{n-1}} = R\ket{\bar{0}}$. In other words, in the expansion of $E$, there is no term corresponding to either $\delta t^0$, $\delta t^1$ or $\delta t^2$. Therefore, one can obtain the inequality~(\ref{eq:E}), where the second and third lines negate the $\delta t^0$, $\delta t^1$ and $\delta t^2$ terms from the first line. We can sum the $\delta t^3$ terms from the first line of inequality~(\ref{eq:E}) as follows,
\begin{eqnarray}
\Delta^{(3)} &=& \norm{H} \norm{H^2} + \frac{1}{3}\norm{H^3} \notag \\
&&+ \norm{\frac{d R}{dt}} \norm{\frac{d^2 R}{dt^2}} + \frac{1}{3}\norm{\frac{d^3 R}{dt^3}} \notag \\
&&+ \norm{H}(\norm{\frac{d R}{dt}}^2 + \norm{\frac{d^2 R}{dt^2}}) \notag \\
&&+ (\norm{H}^2 + \norm{H^2}) \norm{\frac{d R}{dt}}.
\end{eqnarray}

\section{Error Twirling of controlled-phase gates}
\label{sec:twirling}

The twirled controlled-phase gate on a state $\rho$ reads 
\begin{eqnarray}
\frac{1}{16}\sum _{a = 0}^3 \sum _{b = 0}^3
[\SIG{c}{\rm c} \SIG{d}{\rm t}]
\mathcal{N}_\Lambda \mathcal{U}_\Lambda
[\SIG{a}{\rm c} \SIG{b}{\rm t}] \rho
= \bar{\mathcal{N}}_\Lambda \mathcal{U}_\Lambda \rho
\end{eqnarray}
The effective noise superoperator is
\begin{eqnarray}
\bar{\mathcal{N}}_\Lambda &=& \frac{1}{16}\sum _{a = 0}^3 \sum _{b = 0}^3
[\SIG{a}{\rm c} \SIG{b}{\rm t}] \mathcal{N}_\Lambda [\SIG{a}{\rm c} \SIG{b}{\rm t}].
\end{eqnarray}
Using $\SIG{a}{} \SIG{b}{} \SIG{a}{} = [2\delta_{a,b}-(2\delta_{a,0}-1)(2\delta_{b,0}-1)]\SIG{b}{}$, we find that the effective noise is in the stochastic form of Eq.~(\ref{eq:effective}).

\section{Some details about numerical simulations}
\label{sec:DetNum}

In the hybrid algorithm with error reduction, we take $r_1 = 1$ and $r_2 = 2$ and fit the function $\mean{X} = \mean{X}^{(0)} + \chi r$ to infer the value of $\mean{X}^{(0)}$. To simulate shot noise, we use the normal distribution to approximate the binomial distribution, i.e.~the value of $\mean{X}$ taking into account the shot noise is given by $x = 1 - 2p$, where $p$ is given by a normal distribution with the mean $p_0 = (1-\mean{X})/2$ and the standard deviation $\sqrt{p_0(1-p_0)/N_{\rm r}}$.

In the Trotterisation algorithm, the time evolution is simulated using Eq.~(\ref{eq:decomposition}). For the quantum Ising model, the Hamiltonian is decomposed into two terms $H_1 = H_{\rm Z}$ and $H_2 = H_{\rm X}$. In our numerical simulations, we take $N_{\rm t} = {\rm ceiling}(T/\delta t)$, $\tau_{n,j} = \delta t$ if $n<N_{\rm t}$, and $\tau_{N_{\rm t},j} = T - (N_{\rm t}-1)\delta t$. Such a method of determining Trotterisation parameters coincides with the optimal choice of the number $N_{\rm t}$ to minimise errors~\cite{Knee2015}. The state $\rho(t)$ is given by replacing $T$ with $t$ and determining $N_{\rm t}$ and $\tau_{n,j}$ using the method we just described.

To find the optimal $\delta t$, we consider the average trace distance $\overline{D(\ket{\Phi},\rho)} = T^{-1}\int_0^T dt D(\ket{\Phi(t)},\rho(t))$, where $\ket{\Phi}$ is the true state, and $\rho$ is the state prepared in the quantum computer according to the Trotterisation algorithm. The average trace distance is plotted in Fig.~\ref{fig:plot}(a), taking $T = 4\pi$. To obtain other results in Fig.~\ref{fig:plot}, we take $\delta t = 2\pi \times 10^{-1.4}$ in the Trotterisation algorithm, which minimises the average distance.

It is interesting to ask whether second-order techniques, which are known to be helpful in reducing the error in the Trotterisation technique, might have a superior performance here. For the lowest-order symmetric Trotter-Suzuki decomposition~\cite{Berry2007, Wiebe2010}, each time step in Eq.~(\ref{eq:decomposition}) is replaced by $\prod_{j=1}^{N_{\rm H}}e^{-iH_j \delta t/2} \prod_{j=N_{\rm H}}^{1}e^{-iH_j \delta t/2}$, where $N_{\rm H}$ is the number of terms in the Hamiltonian. The performance when using this technique is plotted as the grey curve in the inset to Fig.~\ref{fig:plot}(a). We see that the performance is actually inferior to the more basic Trotterisation approach; the explanation is that the potential gains are more than negated because the larger number of gates employed introduces a greater degree of error (due to the $0.1\%$ physical gate errors).


\begin{thebibliography}{9}
\bibitem{footnote1} For the $(i,j)$ term of $M_{k,q}$, $a = 2\abs{if_{k,i}^*f_{q,j}}$, $\theta = \arg{if_{k,i}^*f_{q,j}}$ and $U = R^\dag_{k,i}R_{q,j}$; for the $(i,j)$ term of $V_{k}$, $a = 2\abs{f^*_{k,i}h_{j}}$, $\theta = \arg{f^*_{k,i}h_{j}}$ and $U = R^\dag_{k,i}\sigma_{j}R$.
\bibitem{footnote2} In the circuit, there are only two controlled gates because of the pattern of $R_l$ terms in $\bra{\bar{0}}U\ket{\bar{0}}$. Note that $U$ is a product of unitary operators, and if $R_l$ is in the product $R_l^\dag$ is also in the product. The only two exceptions are $U_k$ and $U_q$. The circuit prepares the state $(\ket{0}\otimes R_{N_{\rm v}} \cdots R_{k} U_{k} \cdots R_{1} \ket{\bar{0}} + e^{i\theta}\ket{1}\otimes R_{N_{\rm v}} \cdots R_{q} U_{q} \cdots R_{1} \ket{\bar{0}})/\sqrt{2}$ before the ancillary qubit is measured. If we take out an overall factor $R_{N_{\rm v}} \cdots R_{k} U_{k} \cdots R_{1}$, which does not affect the measurement outcome, the state becomes $(\ket{0}\otimes \ket{\bar{0}} + e^{i\theta}\ket{1}\otimes U \ket{\bar{0}})/\sqrt{2}$, which is the state prepared using the circuit in Ref.~\cite{Ekert2002}.
\end{thebibliography}
\end{document}